\documentclass[aps,pra,twocolumn,groupedaddress,showpacs]{revtex4-1}

\usepackage{graphicx}
\usepackage{amsmath}
\usepackage{textcomp}
\usepackage{mathrsfs}
\usepackage{amsfonts}

\begin{document}

\newcommand{\beq}{\begin{equation}}
\newcommand{\eeq}{\end{equation}}
\newcommand{\barr}{\begin{eqnarray}}
\newcommand{\earr}{\end{eqnarray}}
\newcommand{\bseq}{\begin{subequations}}
\newcommand{\eseq}{\end{subequations}}

\newcommand{\vett}[1]{\mathbf{#1}}
\newcommand{\uvett}[1]{\hat{\vett{#1}}}
\newcommand{\mat}[4]{\left[
\begin{array}{cc}
#1 & #2 \\ #3 & #4 \\
\end{array}
\right]}
\newcommand{\barm}{\bar{m}}
\newcommand{\barv}{\tilde{\nu}}

\title{Radially Self-Accelerating Optical Pulses}

\author{Marco Ornigotti$^{1,2}$}

\email{marco.ornigotti@uni-rostock.de}

\author{Alexander Szameit$^1$}

\affiliation{$^1$Institut f\"ur Physik, Universit\"at Rostock, Albert-Einstein-Stra\ss e 23, 18059 Rostock, Germany}
\affiliation{$^2$ Laboratory of Photonics, Tampere University of Technology, P.O. Box 692, FI-33101 Tampere, Finland}

\date{\today}

\begin{abstract}
We generalise the concept of radially self-accelerating beams, to the domain of optical pulses. In particular, we show, how radially self-accelerating optical pulses (RSAPs) can be constructed by suitable superpositions of X-waves, which are a natural extension of Bessel beams in the pulsed domain. Moreover, we show, that while field rotating RSAPs preserve their self-acceleration character, intensity rotating RSAPs only possess pseudo self-acceleration, as their transverse intensity distribution is deformed during propagation, due to their propagation-dependent angular velocity. 
\end{abstract}

\pacs{41.20.Jb, 42.25.Fx, 42.25.Hz}

\maketitle

\section{Introduction}
In recent years, accelerating electromagnetic fields, i.e., solutions of Maxwell's equation, which propagate along curved trajectories in free space, without being subject to an external force, have been the subject of a rather intensive research. The archetype of accelerating beam, is surely the Airy beam, first introduced in quantum mechanics by Berry and Balasz in 1974 \cite{ref1}, and then brought to optics by Siviloglou and co-workers in 2007 \cite{ref2,ref3}. Do to their exotic nature, and novel features, Airy beams were studied within the context of nonlinear optics \cite{ref4}, particle manipulation \cite{ref4bis}, and gave rise to very interesting and innovative applications, such as the generation of curved plasma channels \cite{ref5}. Since 2007, accelerating beams were studied in different coordiante systems \cite{ref6,ref7}, their trajectory was suitably engineered to match different form of curved \cite{ref8,ref9,ref10} and arbitrary \cite{ref11} paths, and find new schemes of acceleration, such as radial \cite{nostroPRL,ref13}, and angular \cite{ref14, ref15} accelerating beams. The former, in particular, are often referred to as radially self-accelerating beams (RSABs), and propagate along spiralling trajectories around their optical axis, due to radial acceleration.

RSABs are typically described, in the monochromatic regime, in terms of superpositions of Bessel beams, with an angular velocity proportional to the amount of orbital angular momentum they carry \cite{nostroPRL}. The distinguishing characteristic of RSABs, however, is a transverse intensity distribution, that rotates around the propagation axis, without exhibiting diffraction, a consequence of RSABs being represented as a sum of nondiffracting beams. RSABs, moreover, have potential applications in different areas of physics, such as sensing \cite{ref5}, material processing \cite{ref16,ref17}, and particle manipulation \cite{ref18,ref19}. Despite this broad interest, however, RSABs have been so far only studied within the monochromatic regime, and the possibility of extending their properties to the domain of optical pulses, has not been investigated yet. Having at hand radially self-accelerating pulses, in fact, could drastically benefit, for example their applications in material processing, or particle manipulation, to name a few. 

In this work, we focus the attention on the generalisation of the concept of self-acceleration to the pulsed domain. In doing that, we will show, how it is possible to create radially self-accelerating pulses (RSAPs) using superpositions of X-waves, rather than Bessel beams. This simple extension of the definition of RSAB given in Ref. \cite{nostroPRL}, however, has some important consequences on the nature of the self-accelerating character of such pulses.

This work is organised as follows: in Sect. II, we briefly recall the properties and definition of RSABs. Then, in Sect. III, we show that RSAPs  can be constructed by suitably generalising their definition in the monochromatic domain, as a superposition of X-waves, rather than Bessel beams, for both the cases of field rotating, and intensity rotating RSAPs. For the latter case, we show, that the only possible analytical form of intensity rotating RSAPs, can be obtained, by assigning a different propagation constant, to each monochromatic beams, composing the pulse Finally, conclusions are drawn in Sect. IV.
\section{Radially Self-Accelerating Beams}
\begin{figure}[!t]
\begin{center}
\includegraphics[width=0.55\textwidth]{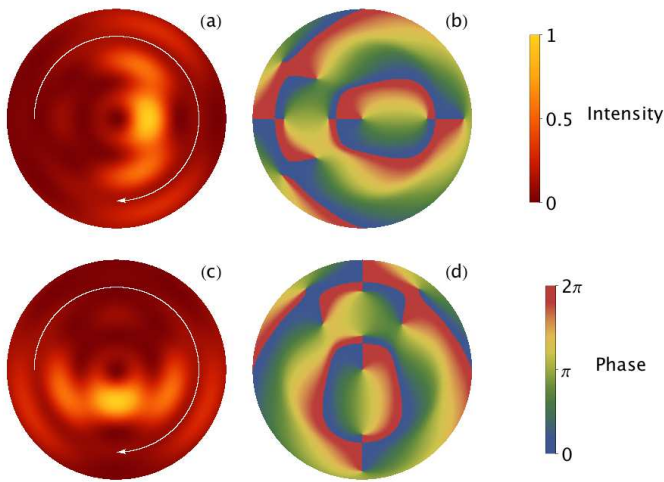}
\caption{Intensity and phase distribution for field rotating RSABs. Panels (a) and (c) correspond to the intensity distributions at $z=0$, and $z=\pi/2\Omega$, respectively, while panels (b) and (d) depict the correspondent phase profiles. The intensity and phase distributions have been plotted in the region $0<R<12$ $\mu m$. For these figures, $\Omega=75$ $rad/s$, $\lambda=800$ $nm$, $\beta=0$ and $C_m=1$ for $0<m\leq 4$, and $C_m=0$, otherwise, have been used. The white arrow in the intensity distribution shows the direction of rotation of the RSAB.}
\label{figure1}
\end{center}
\end{figure}

\begin{figure}[!t]
\begin{center}
\includegraphics[width=0.55\textwidth]{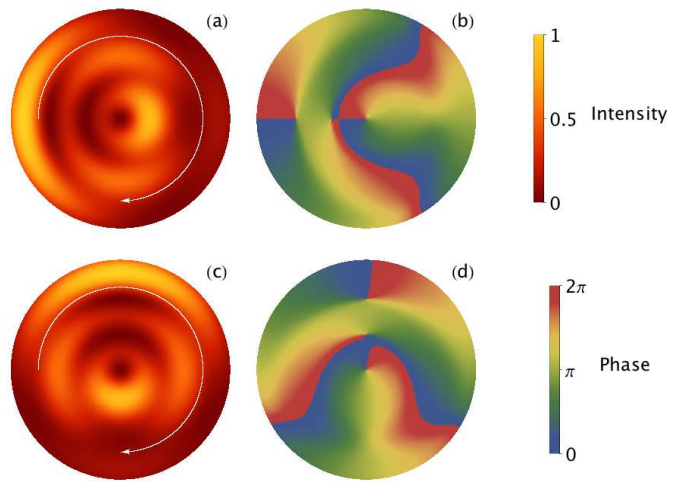}
\caption{Intensity and phase distribution for intensity rotating RSABs. Panels (a) and (c) correspond to the intensity distributions at $z=0$, and $z=\pi/2\Omega$, respectively, while panels (b) and (d) depict the correspondent phase profiles. The intensity and phase distributions have been plotted in the region $0<R<1.5$ $mm$. The difference in plotting range with respect to Fig. \ref{figure1} reflects the paraxial character of intensity rotating RSABs. in contrast to the nonparaxial character of their field rotating counterparts. For these figures, $\Omega=75$ $rad/s$, $\lambda=800$ $nm$, $\beta=7.8$ $\mu m^{-1}$, and $C_m=1$  for $0<m\leq 4$, and $C_m=0$, otherwise, have been used. The white arrow in the intensity distribution shows the direction of rotation of the RSAB.}
\label{figure2}
\end{center}
\end{figure}

As a starting point of our analysis, let us consider a scalar, monochromatic beam, solution of the free space Helmholtz equation
\beq\label{eq1}
\left(\nabla^2+k^2\right)\psi(\vett{r};k)=0,
\eeq
where $k=2\pi/\lambda$ is the vacuum wave vector of the beam, and $\lambda$ its wavelength. In cylindrical coordinates, the most general solution to the above equation can be written in terms of Bessel beams, as follow
\beq\label{eq2}
\psi(\vett{r};k)=\sum_m\,\int\,d\kappa\,A_m(\kappa)\,\text{J}_m\left(R\sqrt{k^2-\kappa^2}\right)e^{im\theta+\kappa z},
\eeq
where $\text{J}_m(x)$ is the Bessel function of the first kind \cite{nist}, and the integration variable $\kappa\propto\cos\vartheta_0$ represents the characteristic Bessel cone angle \cite{durnin}. Following the prescriptions of Ref. \cite{nostroPRL}, is it possible to extract RSABs from the above equation by choosing $A_m(\kappa)=C_m\delta(\kappa-(m\Omega+\beta))$, where $\Omega$ is the actual angular velocity of the RSAB, and $\beta$ is a free parameter, with the dimension of a propagation constant. This choice ensures the possibility to define a co-rotating reference frame $\Phi=\theta+\Omega z$, in which the RSAB appears propagation invariant, namely $\partial\psi_{RSAB}(\vett{r};k)/\partial z=0$. The explicit form of a RSAB thus reads
\beq\label{eq3}
\psi_{RSAB}(R,\Phi;k)=e^{i\beta z}\sum_{m\in\mathcal{M}}C_m\,\text{J}_m(\alpha_m R)e^{im\Phi},
\eeq
where $\alpha_m=\sqrt{k^2-(m\Omega+\beta)^2}$ represents the transverse wave vector of the single Bessel component of the RSAB, and $\mathcal{M}=\{m\in\mathbb{N}: \alpha_m>0\}$. For $\beta=0$, the above equation represents the so-called field rotating RSABs, for which both amplitude and phase spiral around the propagation direction synchronously. For $\beta\neq 0$, instead, Eq. \eqref{eq3} describes the intensity-rotating RSABs, where the amplitude and phase distributions are not synchronised anymore during their rotation along the propagation direction, although the intensity distribution remains propagation invariant. An example of field rotating and intensity rotating RSABs is given in Fig. \ref{figure1} and \ref{figure2}, respectively. 
\section{Extension to Pulse Domain}
To extend the concept of RSABs to the polychromatic domain, we first notice that given a solution $\psi(\vett{r};k)$ of the Helmholtz equation \eqref{eq1}, it is possible to construct an exact solution of the wave equation
\beq\label{eq4}
\left(\nabla^2-\frac{1}{c^2}\frac{\partial^2}{\partial t^2}\right)F(\vett{r},t)=0,
\eeq
as follows:
\beq\label{eq5}
F(\vett{r},t)=\int\,dk\,g(k)\,e^{-ickt}\,\psi(\vett{r};k),
\eeq
where $g(k)$ is an arbitrary spectral function. If we then substitute to $\psi(\vett{r};k)$ the expression of a RSABs, as given by Eq. \eqref{eq3}, we obtain the general expression for a radially self-accelerating pulse (RSAP), namely
\barr\label{eq6}
F_{RSAP}(\vett{r},t)&=&\sum_{m\in\mathcal{M}}C_m\,e^{im\Phi}\,\int\,dk\,g(k)\,e^{i(\beta z-ckt)}\nonumber\\
&\times&\text{J}_m(R\sqrt{k^2-(m\Omega+\beta)^2}).
\earr
Before proceeding any further, it is worth spending a couple of words on the general structure of the above integral. First of all, we can distinguish two different cases, namely $\beta=0$, corresponding to field rotating RSAPs, and $\beta\neq 0$, corresponding to intensity rotating RSAPs. The latter case, however, can be further divided into two sub-classes, namely the case $\beta=\beta(k)$ (meaning, that each monochromatic component of the RSAP defined in Eq. \eqref{eq6} will have its own global propagation constant), and the case $\beta=\text{const}\neq 0$. In the latter case, discussed below in Sect. II.B, the spectrum of the RSAP is  $m$-dependent, meaning, that each component in the sum in Eq. \eqref{eq6} has to first be transformed into a polychromatic signal with its own spectrum, and then summed to form the RSAP. We will show that, the case $\beta=\beta(k)$ results in a pseudo self-accelerating pulse, where self-acceleration is restored only asymptotically, while the case $\beta=\text{const}.$ will instead give rise to a rigorous, self-accelerating pulse.
\subsection{Intensity Rotating RSAPs with $\beta=\beta(k)$}
Let us first consider the case $\beta=\beta(k)\neq 0$. First, we observe, that, typically, $\Omega\ll k$, meaning that the rotation rate of the RSAB is much smaller, than its actual wave vector. If we substitute this Ansatz into the argument of the Bessel function appearing in Eq. \eqref{eq6}, we can Taylor expand the square root appearing as argument of the Bessel function in Eq. \eqref{eq6} with respect to the small parameter $\Omega/k$, thus obtaining
\barr\label{eq7}
&&\sqrt{k^2-(m\Omega+\beta)^2}\simeq k\Bigg[\sqrt{1-\frac{\beta^2}{k^2}}\nonumber\\
&-&\frac{\beta}{\sqrt{k^2-\beta^2}}\left(\frac{m\Omega}{k}\right)-\frac{km^2\Omega^2}{2(k^2-\beta^2)^{3/2}}\nonumber\\
&+&\mathcal{O}\left(\frac{\Omega^3}{k^3}\right)\Bigg].
\earr
Since $\beta=\beta(k)$ can be chosen arbitrarily, we can assume, without loss of generality, that it can be written as $\beta(k)=k\cos\xi$, where $0<\xi<\pi/2$. If we do so, we can simplify the expansion above as follows:
\beq\label{eq7bis}
\sqrt{k^2-(m\Omega+\beta)^2}\simeq k\sin\xi-m\Omega\cot\xi+\mathcal{O}\left(\frac{\Omega^2}{k^2}\right),
\eeq
or, by defining $\Lambda=\Omega(\cos\xi/\sin^2\xi)$ as the new angular velocity of the RSAP, we obtain
\beq\label{eq7ter}
\sqrt{k^2-(m\Omega+\beta)^2}\simeq\sin\xi(k-m\Lambda).
\eeq
This approximation is valid, provided that $(m\Omega)/k\ll1$. Since the number of components of RSABs can be decided almost arbitrarily, however, it is possible to define a new set $\mathcal{M}'=\{m\in\mathbb{N}_0: m\ll(k/\Omega)\}$, and therefore restrict the summation in Eq. \eqref{eq6} to the subset $\mathcal{M}'\subset\mathcal{M}$. If we do so, and introduce the change of variables $k'=k-m\Lambda$, we obtain a rather simple form for RSAPs, namely
\beq\label{eq8}
F_{RSAP}^{(1)}(\vett{r},t)=\sum_{m\in\mathcal{M}'}\,C_me^{im\Theta_0}X_m(R,\zeta),
\eeq
where $\Theta_0=\theta+\Lambda\zeta$ is the co-rotating coordinate, and
\beq\label{eq12}
X_m^{(1)}(R,\zeta)=\int\,dk\,g(k)\,e^{ik\zeta}\text{J}_m(kR\sin\xi),
\eeq 
represents the general expression of a X-wave \cite{localisedWaves, XwavesPRL}, with $\zeta=z\cos\xi-ct$ being its correspondent co-moving coordinate. This is the first result of our work. In the polychromatic domain, radially, self-accelerating fields can be constructed by taking superpositions of X-waves, rather than Bessel beams. 

However, as it can be seen from Eq. \eqref{eq8}, intensity rotating RSAPs intrinsically contain a $\zeta$-dependence on both their co-rotating coordiante $\Theta_0$, and transverse distribution $X_m(R,\zeta)$. This fact, which will be discussed in detail in the next section, is ultimately the reason, why RSAPs only possess a pseudo self-accelerating character.

At a first glance, Eq. \eqref{eq12} has the same form of its monochromatic counterpart, namely Eq. \eqref{eq3}, and could be interpreted as its straightforward generalisation. One, in fact, could naively substitute Bessel beams, which are used in the monochromatic case to generate RSABs, with X-waves (i.e., polychromatic Bessel beams), thus realising RSAPs.

A closer analysis of Eq. \eqref{eq8}, however, reveals an important difference between the two cases, namely that while RSABs describe spiralling trajectories of constant transverse dimension \cite{nostroPRL}, RSAP describe spiralling trajectories of growing transverse dimension. Moreover, while the transverse structure of RSABs rigidly rotates around the propagation axis, this is not the case for RSAPs, which instead show a progressive self-adaption of the transverse intensity distribution to a ring, centered on the propagation axis.

To better understand this, let us consider explicitly the case of fundamental X-waves. These are characterised by an exponentially decaying spectrum, i.e., $g(k)=\text{H}(k)\exp{[-\alpha k]}$, where $\alpha$ accounts for the width of the spectrum, and has the dimensions of a length, and $\text{H}(x)$ is the Heaviside step function. If we substitute this exponentially decaying spectrum into Eq. \eqref{eq8}, and use Eq. 6.621.1 in Ref. \cite{gradsteyn}, we get, after some simple algebraic manipulation, the following result:
\begin{widetext}
\beq\label{eq9}
F_{RSAP}^{(1)}(\vett{r},t)=e^{i\arctan\left(\frac{\zeta}{\alpha}\right)}\sum_{m\in\mathcal{M}'}A_me^{im\Theta}\frac{\rho^m}{\sqrt{\alpha^2+\zeta^2}}\,_2F_1\left(\frac{m+1}{2},\frac{m+2}{2};m+1;-\rho^2e^{2i\arctan\left(\frac{\zeta}{\alpha}\right)}\right),
\eeq
\end{widetext}
where 
\beq
\rho\equiv\rho(\zeta)=\frac{R\sin\xi}{\sqrt{\alpha^2+\zeta^2}},
\eeq
is an expanding, normalised, radial coordinate, 
\beq
\Theta=\theta+\Lambda\zeta+\arctan\left(\frac{\zeta}{\alpha}\right),
\eeq
is the co-rotating, accelerating reference frame, $A_m=C_m/2^m$, and $\,_2F_1(a,b;c;x)$ is the Gauss hypergeometric function \cite{nist}. Notice, that although in general the hypergeometric function gives an extra $m-$ and $\zeta$-dependent phase contribution, which modifies the definition of $\Theta$, if we limit ourselves to the case $\xi\ll 1$, the phase contribution of the hypergeometric function can, at the leading order in $\xi$, be neglected. 

We can now compare the two co-rotating coordiantes, in the monochromatic ($\Phi$) and polychromatic ($\Theta$) case: while $\Phi$ essentially describes an helix centered around the $z$-axis, whose transverse width remains constant, since the angular velocity of the RSAB is $\Lambda=const.$, this is not the case for the polychromatic co-rotating coordinate $\Theta$, as it represents an accelerating coordinate, with velocity
\beq\label{velocity}
\frac{\partial\Theta}{\partial\zeta}=\Lambda+\frac{\alpha}{\alpha^2+\zeta^2},
\eeq
and acceleration 
\beq\label{acceleration}
\frac{\partial^2\Theta}{\partial\zeta^2}=-\frac{2\alpha\zeta}{(\alpha^2+\zeta^2)^2}.
\eeq

The above expressions for the angular velocity and acceleration of the RSAP, reveals that for large enough propagation distances, $\partial\Theta/\partial\zeta\rightarrow\Lambda$, and $\partial^2\Theta/\partial\zeta^2\rightarrow 0$, and the standard values of velocity and acceleration for RSABs are restored. This means, that the self-accelerating state represents an asymptotic equilibrium for the RSAP. For small propagation distances, on the other hand, the behaviour of RSAPs change significantly from traditionally self-accelerating beams, as can be seen from Eqs. \eqref{velocity} and \eqref{acceleration}. Since the co-rotating coordinate is now accelerating, and the acceleration is towards the center of the pulse, the transverse field distribution needs to adapt to this attractive force, which tends (asymptotically) to transform the intensity distribution into a ring-shaped pulse, around the co-moving propagation direction $\zeta$. For these reasons, one cannot formally speak anymore of self-accelerating pulses, as there exists no reference frame, in which Eq. \eqref{eq9} appears propagation invariant, or, said in other terms, there exist no reference frame, in which the motion of the pulse around the $\zeta$-axis, can be described by an helix. However, since the self-accelerating behaviour is an asymptotical equilibrium of the system, one could refer to such pulses as pseudo self-accelerating.

\begin{figure}[!t]
\begin{center}
\includegraphics[width=0.55\textwidth]{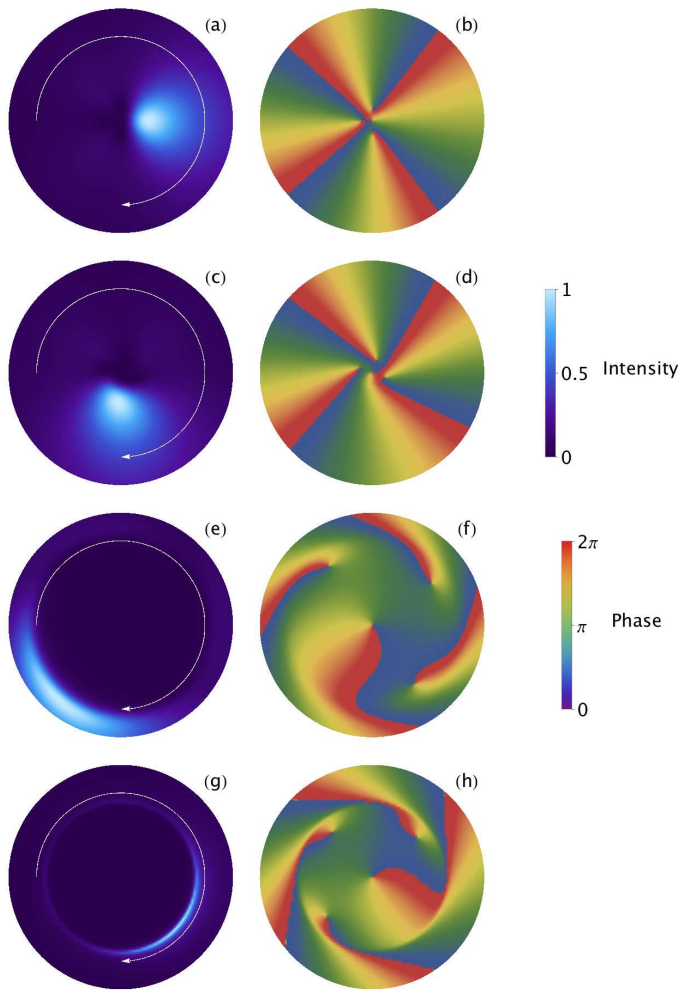}
\caption{Intensity (left) and phase (right) distribution for intensity rotating RSAPs, as defined by Eq. eqref{eq9}. The plots are made at different values of the normalised propagation length $\zeta\equiv\zeta/\alpha$, namely $\zeta=0$ [panels (a), and (b)], $\zeta=2\pi/\Lambda$ [panels (c), and (d)], $\zeta=10\,(2\pi/\Lambda)$ [panels (e), and (f)], and $\zeta=50\,(2\pi/\Lambda)$ [panels (g), and (h)]. As it can be seen, the transverse profile of intensity rotating RSAPs gets progressively distorted, up to the point, at which it stabilises in a ring-shaped form [panel (e)]. The intensity and phase distributions have been plotted in the region $0<\rho<1250$, for the panels (a)-(f), and $0<\rho<7500$, for panels (g), and (h). For these figures, $\Omega=75$ rad/s, $\lambda=800$ nm, $\xi=0.01$ (correspondent to $\beta=7.853\times 10^6$ $m^{-1}$), and $C_m=1$ for $1<m\le 4$, and $C_m=0$, otherwise, have been used. The white arrow in the intensity distribution shows the direction of rotation of the RSAP.}
\label{figure3}
\end{center}
\end{figure}

The intensity and phase distributions of intensity rotating RSAPs are shown in Figs. \ref{figure3}. For small propagation distances [Figs. \ref{figure3}(a)-(d)], the intensity distribution gets progressively distorted, while propagating along $\zeta$, up to the point, in which the RSAP reaches its equilibrium form of a ring [Fig. \ref{figure3}(e), and (f)]. From this point on, the transverse intensity distribution does not change anymore in shape, but only becomes bigger, due to the expanding nature of the radial co-moving coordinate $\rho$, as can be seen by coparing panels (e) and (g) of Fig. \ref{figure3}. If we compare the behavior of RSAP at large $\zeta$, with the one of RSABs, we can notice, that while the transverse dimension of the spiral described by RSABs remains constant (essentially, because $\Phi$ describes a helix, rather than a spiral), this is not the case for RSAPs, since $\Theta$ describes a spiral, whose transverse dimension is growing with $\zeta$. 

To estimate this, let us calculate the average transverse size of the spiral, by considering the position of the center of mass of the RSAP intensity distribution, as follows:
\barr
\langle R(\zeta)\rangle&=&\int\,d^2R\,R\,\left|F_{RSAP}(\vett{r},t)\right|^2\nonumber\\
&\equiv&R_0\sqrt{\alpha^2+\zeta^2},
\earr
where, at the leading order in $\xi$ \cite{note2},
\beq\label{erre0}
R_0=\frac{2\pi}{\sin^3\xi}\sum_{m\in\mathcal{M'}}|A_m|^2\int_0^{\infty}d\rho\,\rho^{2(m+1)},
\eeq
Thus, in the co-moving, expanding, reference frame, the transverse dimension of the spiral grows as $\sqrt{\alpha^2+\zeta^2}$.
\subsection{Intensity Rotating RSAPs with $\beta=\text{const}$}
Another possibility for intensity rotating RSAPs, is to choose $\beta=\text{const}.\neq 0$. If we use this assumption, introduce the change of variables $k'=\sqrt{k^2-(m\Omega+\beta)^2}$ in Eq. \eqref{eq6}, we allow the spectral function $g(k)$ to be $m$-dependent, and we redefine it as $g_m(k)=\left(2kG_m(k)/\sqrt{k^2-(m\Omega+\beta)^2}\right)\text{H}(k)$, where $\text{H}(k)$ is the Heaviside step function \cite{nist}, we get the following result
\beq\label{eqNew}
F_{RSAP}^{(2)}(\vett{r},t)=e^{i\beta z}\sum_{m\in\mathcal{M}'}C_me^{im\Phi}\, X_m^{(2)}(R,t;\beta),
\eeq
where
\beq\label{effeDue}
X_m^{(2)}(\vett{r};\beta)=\int_0^{\infty}dk\,G_m(k)e^{-ict A(k)}\text{J}_m(kR),
\eeq
where $A(k)=\sqrt{k^2+(m\Omega+\beta)^2}$. If we choose the spectral function $G(k)$ as
\beq
G(k)=\frac{1}{\sqrt{k^2+(m\Omega+\beta)^2}},
\eeq
Eq. \eqref{effeDue} admits the following, closed form analytical solution \cite{gradsteyn}
\barr
X_m^{(2)}(R,t;\beta)&=&I_{m/2}\left(\frac{\alpha}{2}\left(\sqrt{R^2-c^2t^2}-ict\right)\right)\nonumber\\
&\times&K_{m/2}\left(\frac{\alpha}{2}\left(\sqrt{R^2-c^2t^2}+ict\right)\right),
\earr
where $I_m(x)$, and $K_m(x)$ are the modified Bessel function of the first, and second kind, respectively \cite{nist}, and $\alpha=m\Omega+\beta$.

As it can be seen, no $z$-dependence is contained in the expression of the transverse field distribution $X_m^{(2)}(R,t;\beta)$, and Eq. \eqref{eqNew} has the same form of Eq. \eqref{eq3}. In this case, therefore, we can define the co-rotating reference frame $\{R,\Phi,t\}$, where $F_{RSAP}^{(2)}(\vett{r},t)$ is manifestly propagation invariant. However, due to the presence of the modified Bessel function of the second kind $K_{m/2}(x)$, this field distribution is divergent in the origin, and therefore it cannot represent a physically meaningful solution. Analytical forms of intensity rotating RSAPs, therefore, only exist for $\beta=\beta(k)$. This is the second result of our work: intensity rotating RSAPs can be only constructed by assigning a proper $\beta(k)$ to each of the monochromatic components, that contribute to the pulse. If $\beta$ is constant, no physically meaningful analytical solution can be found. 

Notice, however, that a trivial possibility for describing intensity rotating RSAPs with constant $\beta$, would be to consider the case $\beta=\text{const.}\ll 1$. This, however, can be treated (at least at the leading order in $\beta$), as a first order correction to the case $\beta=0$, which will be discussed in the next section. Since $\beta=0$ will correspond to field rotating RSAPs, one could then conclude, that intensity rotating RSAPs with constant small $\beta$, can be well described by field rotating RSAPs.
\subsection{Field Rotating RSAPs}
\begin{figure}[!t]
\begin{center}
\includegraphics[width=0.55\textwidth]{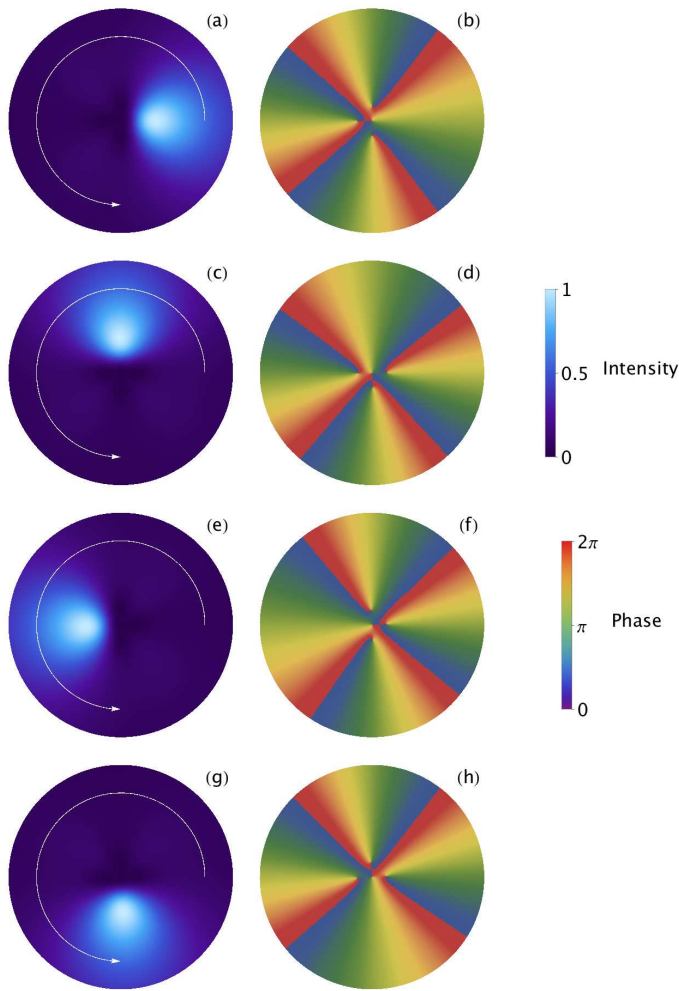}
\caption{Intensity (left) and phase (right) distribution for field rotating RSAPs, at different values of the propagation length namely $\zeta=0$ [panels (a), and (b)], $\zeta=0.25(2\pi/\Lambda)$ [panels (c), and (d)], $\zeta=0.5\,(2\pi/\Lambda)$ [panels (e), and (f)], and $\zeta=0.75\,(2\pi/\Lambda)$ [panels (g), and (h)]. As it can be seen, the transverse profile of field rotating RSAPs remains propagation invariant, and rotates synchronously with its phase profile. The intensity and phase distributions have been plotted in the region $0<R<10$ $\mu m$. For these figures, $\Omega=75$ rad/s, $\lambda=800$ nm, and $C_m=1$ for $1<m\le 4$, and $C_m=0$, otherwise, have been used. The white arrow in the intensity distribution shows the direction of rotation of the RSAP.}
\label{figure4}
\end{center}
\end{figure}

For the case $\beta=0$, the argument fo the Bessel function in Eq. \eqref{eq6} becomes $\sqrt{k^2+m^2\Omega^2}$. If we then assume that $\Omega\ll k$, and use the expansion in Eq. \eqref{eq7} with $\beta=0$,  up to $\mathcal{O}(\Omega^2/k^2)$, we can write $\sqrt{k^2-m^2\Omega^2}\simeq k$. This approximation, however, holds as long as $m$ is chosen in such a way, that $m\Omega/k\ll1$. As stated before, since we are free to choose the set in which $m$ is defined, we can restrict the original set $\mathcal{M}$ to the new subset $\mathcal{M}''\equiv\{m\in\mathbb{N}_0: m\ll (\sqrt{2}k/\Omega)\}$. In this case, the explicit expression of field rotating RSAPs is given as follows:

\beq\label{fieldRSAP}
F_{RSAP}^{(3)}(\vett{r},t)=\sum_{m\in\mathcal{M}''}C_me^{im\Phi}X_m^{(3)}(R,\zeta_0),
\eeq
where $\zeta_0=-ct$, and
\beq
X_m^{(3)}(R,\zeta_0)=\int\,dk\,g(k)\,e^{ik\zeta_0}\text{J}_m(kR).
\eeq
Notice, that unlike the case $\beta\neq 0$, no $z$-dependence is present in the transverse form of the pulse $X_m^{(3)}(R,\zeta_0)$. This means that, field rotating RSAPs are truly self-accelerating fields, as one can define a reference frame, namely $\{R,\Phi,\zeta_0\}\equiv\{R,\Phi,t\}$, in which the RSAP appears propagation invariant, and fulfils all the required conditions for self-acceleration \cite{nostroPRL}. The intensity and phase distributions for field rotating RSAPs at different propagation lengths, are shown in Fig. \ref{figure4}. As it can be seen, the transverse intensity profile remains unchanged, as the pulse propagates along $z$. Notice, moreover, that field rotating RSAPs rotate in the opposite direction, with respect to intensity rotating RSAPs. 
\section{Conclusions}
In this work, we have generalised the concept of radially self-accelerating field, to the domain of optical pulses. We have shown, how it is possible to define RSAPs as a superposition of OAM-carrying X-waves, rather than Bessel beams. For the case of fundamental X-waves, we have calculated the explicit expression for field rotating, as well as intensity rotating RSAPS, and we have shown, that while the former retains their self-acceleration character, the latter possess pseudo self-acceleration, and admit pure self-acceleration only asymptotically. We have also investigated intensity rotating RSAPs with constant $\beta$, and shown, that although in this case it is possible to recover pure self-acceleration, such field bare no physical meaning, as they are divergent in the origin. Our work represents the first attempt, to generalise the concept of self-acceleration to the domain of optical pulses, and discusses advantages and limitations of this process. Moreover, our work represents a guideline, which will be useful for the experimental realisation of radially self-accelerating optical pulses. 
\section*{Acknowledgements}
The authors wish to thank the Deutsche Forschungsgemeinschaft (grant SZ 276/17-1) for financial support.


\begin{thebibliography}{999}

\bibitem{ref1}  M. V. Berry and  N. L. Balazs, \emph{Am, J. Phys.} \textbf{47}, 274 (1979).
%
\bibitem{ref2} G. A. Siviloglou, J. Broky, A. Dogariu, and D. N. Christodoulides, \emph{Phys. Rev. Lett.} \textbf{99}, 213901 (2007).
%
\bibitem{ref3} G. A. Siviloglou, and D. N. Christodoulides, \emph{Opt. Lett.} \textbf{32}, 979 (2007).
%
\bibitem{ref4} Y. Hu, G. A. Siviloglou, P. Zhang, N. K. Efremedis, D. N. Christodoulides, and Z. Chen Z, \emph{Self-accelerating Airy Beams: Generation, Control, and Applications} in \emph{Nonlinear Photonics and Novel Optical Phenomena} (Berlin, Germany: Springer, 2012).
%
\bibitem{ref4bis}J. Baumgartl, M. Mazilu, and K. Dholakia, \emph{Nat. Photonics} \textbf{2}, 675 (2008).
%
\bibitem{ref5} P. Polynkin, M. Kolesik, J. V. Moloney, G. A. Siviloglou, D. N. Christodoulides, \emph{Science} \textbf{10}, 5924 (2009).
%
\bibitem{ref6} M. A. Bandres, \emph{Opt. Lett.} \textbf{33}, 1678 (2008).
%
\bibitem{ref7} M. A. Bandres, \emph{New J. Phys.} \textbf{15}, 013054 (2013).
%
\bibitem{ref8} M. A. Bandres, M. A. Alonso, I. Kaminer, and M. Segev, \emph{Opt. Express} \textbf{21}, 13917 (2013).
%
\bibitem{ref9} M. A. Bandres, \emph{Opt. Lett.} \textbf{34}, 3791 (2009).
%
\bibitem{ref10} I. Kaminer, R. Bekenstein, J. Nemirovsky, and M. Segev, \emph{Phys. Rev. Lett.} \textbf{108}, 163901 (2012).
%
\bibitem{ref11} J. Zhao, I. D. Chremmos, D. Song, D. N. Christodoulides, N. K. Efremidis, and Z. Chen, \emph{Scientific Reports} \textbf{5}, 12086 (2015).
%
\bibitem{nostroPRL} C. Vetter, T. Eichelkraut, M. Ornigotti, and A. Szameit, \emph{Phys. Rev. Lett.} \textbf{113}, 183901 (2014).
%
\bibitem{ref13}C. Vetter, T. Eichelkraut, M. Ornigotti, and A. Szameit, \emph{Appl. Phys. Lett.} \textbf{107}, 211104 (2015).
%
\bibitem{ref14} J. Webster, C. Rosales-Guzman, and A. Forbes, \emph{Opt. Lett.} \textbf{42}, 675 (2017).
%
\bibitem{ref15} C. Vetter, A. Dudley, A. Szameit, and A. Forbes, \emph{Opt. Express} \textbf{25}, 20530 (2017).
%
\bibitem{ref16} A. Jesacher, and M. J. Booth, \emph{Opt. Express} \textbf{18}, 21090 (2010).
%
\bibitem{ref17} A. Mathis, L. Froehly, L. Furfaro, M. Jacquot, J. M. Dudley, and F. Courvoisier, \emph{J. Eur. Opt. Soc.: Rapid Publ.} \textbf{8}, 13019 (2013).
%
\bibitem{ref18} D. McGloin, V. Garces-Chavez, and K. Dholakia, \emph{Opt. Lett.} \textbf{28}, 657 (2003).
%
\bibitem{ref19} S. Sukhov, and A. Dogariu, \emph{Opt. Lett.} \textbf{35}, 3847 (2010).
%
\bibitem{nist} F. W. J. Olver, D. W. Lozier, R. F. Boisvert, and C. W. Clark,
NIST Handbook of Mathematical Functions (Cambridge:
Cambridge University Press, 2010)
%
\bibitem{durnin} J. Durnin, J. J. Miceli, and J. H. Eberly, {\it  Phys. Rev. Lett} {\bf 58}, 1499 (1987).
%
\bibitem{localisedWaves} H. E. Hernandez-Figueroa, M. Zamboni-Rached, and E. Recami (editors), \emph{Localized Waves} (Hoboken, NJ: Wiley, 2008).
%
\bibitem{XwavesPRL} M. Ornigotti, C. Conti, and A. Szameit, {\it Phys. Rev. Lett.} {\bf 115}, 100401 (2015).
%
\bibitem{gradsteyn} I. S. Gradshteyn, and I. M. Ryzhiz, \emph{Table of Integrals, Series and Products} (Cambridge, MA: Academic Press, 2006).
%
\bibitem{note2} Please note, that the integral in the definition of $R_0$ in Eq. \eqref{erre0} diverges. This, however, is only due to the particular choice of spectrum $g(k)$, which leads to infinite-energy-carrying X-waves. For a more realistic, and experimentally realisable case, a Gaussian profile for $g(k)$ can be chosen, which corresponds to a finite radial integral in the definition of $R_0$.

\end{thebibliography}
\end{document}